\documentclass[12pt,preprint]{aastex}

\usepackage{graphicx}
\usepackage{amsmath}
\usepackage{natbib}
\newcommand{\pan}{{\tt Pandurata}\, }

\shorttitle{EM Chirps from NS/MH Mergers}
\shortauthors{Schnittman et al.}

\begin{document}

\title{Electromagnetic Chirps from Neutron Star-Black Hole Mergers}

\author{Jeremy D.\ Schnittman}
\affil{NASA Goddard Space Flight Center, Greenbelt, MD 20771}
\affil{Joint Space-Science Institute, College Park, MD 20742}
\email{jeremy.schnittman@nasa.gov}

\author{Tito Dal Canton}
\affil{NASA Postdoctoral Program Fellow, Goddard Space Flight Center,
  Greenbelt, MD 20771}

\author{Jordan Camp}
\affil{NASA Goddard Space Flight Center, Greenbelt, MD 20771}

\author{David Tsang}
\affil{Department of Astronomy, University of Maryland, College Park,
  MD 20742}
\affil{Joint Space-Science Institute, College Park, MD 20742}
\and
\author{Bernard J.\ Kelly}
\affil{CRESST, NASA Goddard Space Flight Center, Greenbelt, MD 20771}
\affil{Department of Physics, University of Maryland Baltimore
  County, Baltimore, MD 21250}
\affil{Joint Space-Science Institute, College Park, MD 20742}

\begin{abstract}
We calculate the electromagnetic signal of a gamma-ray flare coming
from the surface of a neutron star shortly before merger with a black
hole companion. Using a new version of the Monte Carlo radiation
transport code \pan that incorporates dynamic spacetimes, we integrate
photon geodesics from the neutron star surface until they reach a
distant observer or are captured by the black hole. The gamma-ray
light curve is modulated by a number of relativistic effects,
including Doppler beaming and gravitational lensing. Because the
photons originate from the inspiraling neutron star, the light curve
closely resembles the corresponding gravitational waveform: a chirp
signal characterized by a steadily increasing frequency and
amplitude. We propose to search for these electromagnetic chirps
using matched filtering algorithms similar to those used in 
LIGO data analysis.
\end{abstract}


\section{INTRODUCTION}\label{section:intro}

The tremendous excitement generated by the recent LIGO discoveries of
gravitational waves (GWs) from merging binary black holes
\citep{LIGO1,LIGO2} will only be
surpassed by the unambiguous simultaneous detection of an
electromagnetic (EM) counterpart to such a GW signal. With an EM
counterpart, the sky localization of the GW source will likely be
greatly improved, and if a host galaxy can be identified, the
event can be placed in the proper astrophysical context. The galaxy's
redshift could also be used in conjunction with the luminosity
distance---measured independently with the GW signal---to place new
and unbiased constraints on cosmological parameters \citep{Nissanke:2010}. 

Another major advantage of detecting an EM counterpart is that it
could give us the ability to combine with sub-threshold GW triggers
and improve the statistical significance of marginal signals,
potentially lowering the false alarm rate. It is also conceivable that
several weak coincident EM and GW signals could reveal a
population of sources which are individually too weak to be detectable 
by blind GW or EM searches.

One of the most 
promising mechanisms for a LIGO counterpart is a short gamma-ray burst
(SGRB) produced by seeing a binary neutron star (NS/NS) or neutron
star-black hole (NS/BH) binary merger \citep{Eichler:1989}. The disrupted NS will form a
massive, highly magnetized accretion disk around the companion BH,
driving a relativistic jet which will be manifest as a SGRB to
observers oriented along the jet axis. 
For a sub-sample of SGRBs, a precursor gamma-ray flare can be seen
roughly $1-10$ seconds before the peak of the GRB \citep{Troja:2010}. For a nominal NS/BH
binary with masses $1.4M_\odot$ and $10M_\odot$, this would correspond
to hundreds of orbits before merger, with binary separations of
$20-30$ gravitational radii. 

In this paper, we explore the possibility that these precursor flares
come from some emission process on the surface of the neutron star as
it spirals towards the companion black hole. If
so, the resulting gamma-rays will experience the extreme gravitational
forces that govern this highly relativistic and dynamical
system. These effects will be imprinted on the gamma-ray signal that
ultimately reaches a distant observer. We have identified two primary
features in these light curves: special relativistic Doppler beaming,
and magnification due to gravitational lensing. As we will show below,
the detailed properties of the light curve provide information about
the binary separation, inclination angle, and orbital period, 
giving important information complementary to the GW signal.

The most robust feature of this precursor EM light curve is that is
should be locked in phase with the GW signal, also chirping through
increasing frequency and amplitude as the system approaches
merger. Thus we refer to this particular class of counterparts as
``electromagnetic chirps.'' We fully expect that data analysis tools
similar to those used in GW searches may be
fruitful in discovering and interpreting such signals in otherwise
noisy data from gamma-ray observatories such as the Gamma-ray Burst
Monitor (GBM) on Fermi (Dal Canton et al.\ 2017).
We remain intentionally agnostic about the specific physical mechanism
that produces the gamma-ray flash on the NS surface, but one promising
model is that of resonant shattering flares that occur as the binary
orbital frequency sweeps through the eigenfrequencies of the NS normal
modes during the moments leading up to merger
\citep{Tsang:2012,Tsang:2013}. When the quadrupolar crust-core
interface mode is resonantly excited by tidal interaction, a huge
amount of energy can be deposited into the crust, and ultimately
released as gamma-rays. 

\section{LIGHT CURVES IN BINARY SPACETIME}\label{section:light_curves}

We calculate the light curves and spectra from the NS/BH system with
the Monte Carlo radiation transport code \pan
\citep{Schnittman:2013}. To date, \pan has only been applied
to problems with stationary Kerr spacetimes. In order to use it
for the highly dynamic spacetime of a merging compact binary system,
significant modifications were required. First and foremost, we needed
to move from the Hamiltonian formalism described in
\citet{Schnittman:2013}, appropriate for a system with multiple
integrals of motion, to a more generalized Lagrangian approach to
solving the geodesic equation for an arbitrary metric and connection,
better suited for the binary spacetime. 

At the same time, we desired a spacetime formulation that would be
computationally efficient for integrating millions of geodesic
trajectories. Thus, instead of using full numerical relativity data
(typically only available for tens of binary orbits),
we instead opted for a relatively simple analytic description of
the binary spacetime based on a post-Newtonian (PN) approximation to the
orbit \citep{Kelly:2007}.

The binary is completely described by two non-spinning point masses with
$m_1+m_2=M$ and $m_1 \ge m_2$, moving on a circular orbit with binary
separation $D\equiv (GM/c^2)x$ (unless explicitly stated otherwise, we
hereby assume units with $G=c=1$). The angular velocity is given by the 2PN
expression 
\begin{equation}\label{eqn:Omega}
\Omega =
\left[64\frac{x^3}{(1+2x)^6}+\eta\frac{1}{x^4}+\left(-\frac{5}{8\eta}+\eta^2\right)\frac{1}{x^5}\right]^{1/2}M^{-1}\, ,
\end{equation}
with $\eta \equiv m_1 m_2/M^2$. 

The orbit is assumed to be instantaneously circular, but we do evolve
the separation according to the 2.5PN leading quadrupole radiation
reaction terms derived in \citet{Peters:1964}:
\begin{equation}\label{eqn:Peters}
\frac{dx}{dt} = -\frac{64}{5}\eta\frac{1}{x^3}\, .
\end{equation}

The metric is expressed by the standard 3+1 lapse/shift formalism:
\begin{equation}
g_{\mu \nu} = \begin{pmatrix}
-\alpha^2 + \beta^2 & \beta_j \\
\beta_i & \gamma_{ij}\\
\end{pmatrix}\, .
\end{equation}
Following \citet{Campanelli:2006}, we use $\alpha = 2/(1+\psi^4)$,
$\beta_j =0$, and $\gamma_{ij}= \delta_{ij}\psi^4$. The conformal
factor $\psi$ is given by
\begin{equation}
\psi = 1+ \frac{m_1}{2r_1}+\frac{m_2}{2r_2}\, ,
\end{equation}
with $r_1$ and $r_2$ being the simple Cartesian distances between the
spatial coordinate and the primary/secondary masses. For the
Christoffel components $\Gamma^{\rho}_{\mu \nu}$ we take 
the spatial and temporal metric derivatives analytically based
on the trajectory as given by equation (\ref{eqn:Omega}). 

To generate light curves, \pan employs a Monte Carlo ray-tracing
scheme that is based on shooting a large number of photon packets from
the emission region to potential observers at infinity \citep{Schnittman:2013}.
For the NS/BH problem investigated here, we use a simple thermal, optically
thick emission model. Each photon is
launched from the surface of the NS with radius 10 km, isotropic in the local
frame of the star and with a limb-darkening factor appropriate for an
optically thick emitter. This gives the photon's initial position
$\mathbf{x}$ and four-velocity $\mathbf{u}$. The photon is then
propagated forward along the affine parameter $\lambda$ according to
the standard geodesic formula 
\begin{equation}
\frac{d^2x^\rho}{d\lambda^2} = -\Gamma_{\mu \nu}^{\rho}
\frac{dx^\mu}{d\lambda} \frac{dx^{\nu}}{d\lambda}\, .
\end{equation}

When a photon packet reaches a simulated detector surface at some
large radius, its 
direction specifies the appropriate pixel in the image plane, exactly
like a pinhole camera. Since each photon is also tagged with a time
stamp, a movie can be built up over time. Generally, each observer is
located at a specific sky position $(\theta,\phi)$, so the vast
majority of the Monte Carlo photons never contribute to the image. For
circular orbits, we are able to take advantage of the periodic motion
of the binary to efficiently map the azimuthal coordinate to the time
coordinate, so a single movie is produced by combining images from
observers at all azimuthal positions. Because we are also interested
in generating light curves for multiple latitudinal inclination
angles, every photon that escapes the system
ultimately contributes to some observer's light curve, restoring
a remarkable level of efficiency for the Monte Carlo approach. 

\begin{figure}[h]
\caption{\label{fig:movie_stills} Snapshots of thermal emission from
  the surface of a $1.4 M_\odot$ neutron star orbiting a $10 M_\odot$ black
  hole. The observer is at inclination $90^\circ$, edge-on to the
  orbital plane, and the binary separation is $D=10M$. In panel (a)
  the observer, black hole, and neutron star are co-linear, resulting
  in an Einstein ring around the BH, and producing the peak
  magnification; in (b) the NS is moving towards the observer
  with maximum blueshift; (c) shows the strong gravitational lensing
  of photons that are deflected $180^\circ$ by the BH; and (d) is the
  point of maximum redshift.}
\begin{center}
\includegraphics[width=0.35\textwidth]{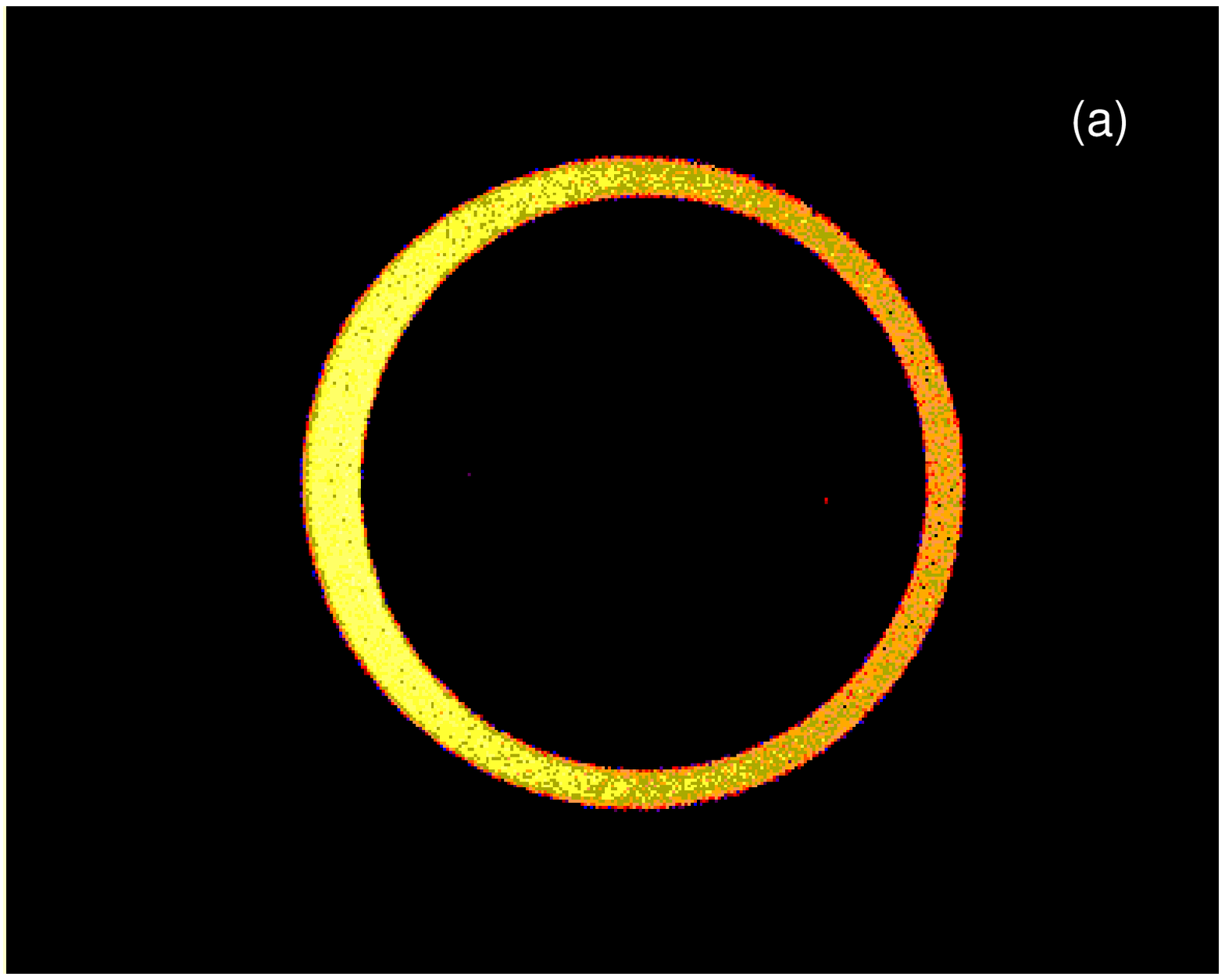}
\includegraphics[width=0.35\textwidth]{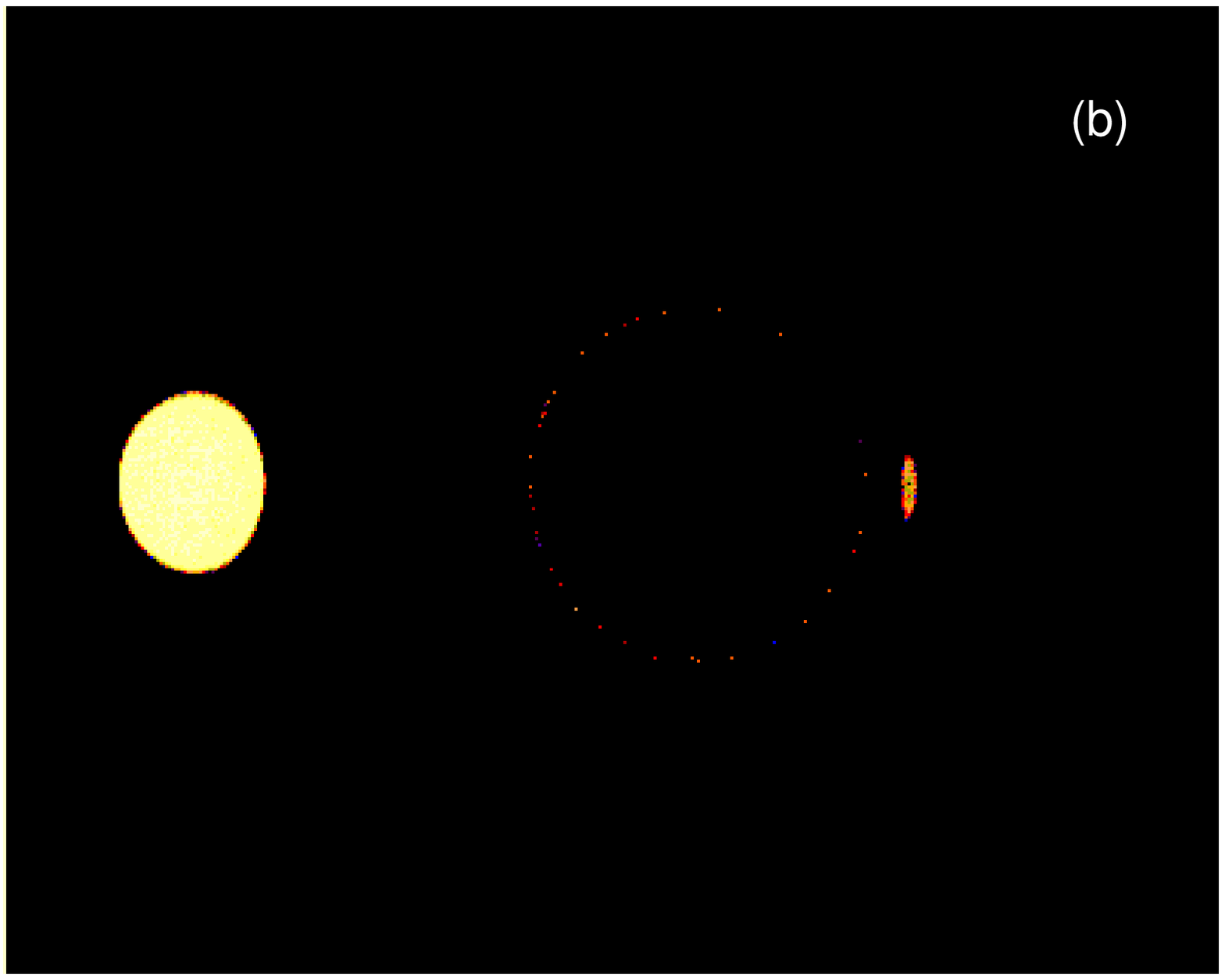}\\
\includegraphics[width=0.35\textwidth]{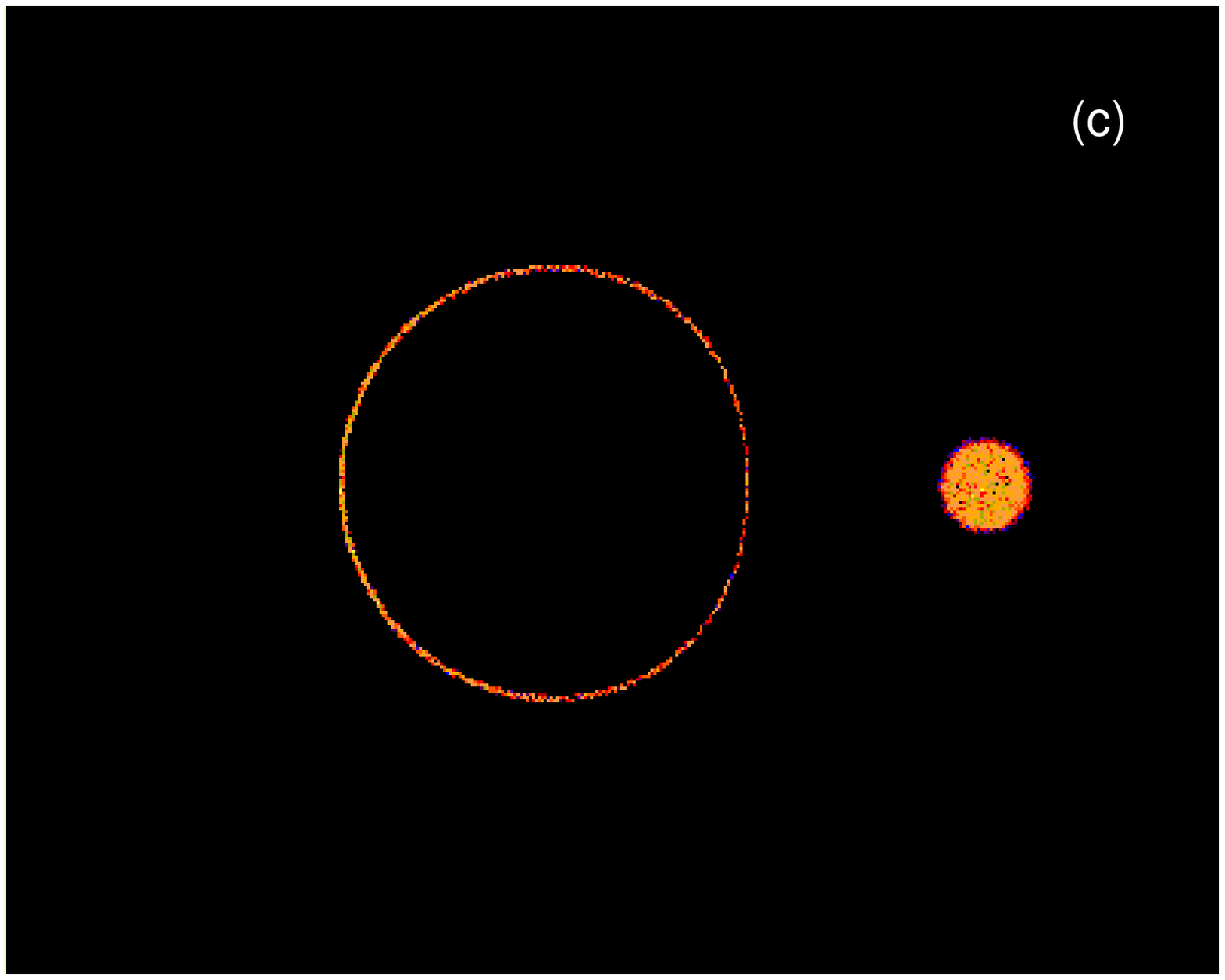}
\includegraphics[width=0.35\textwidth]{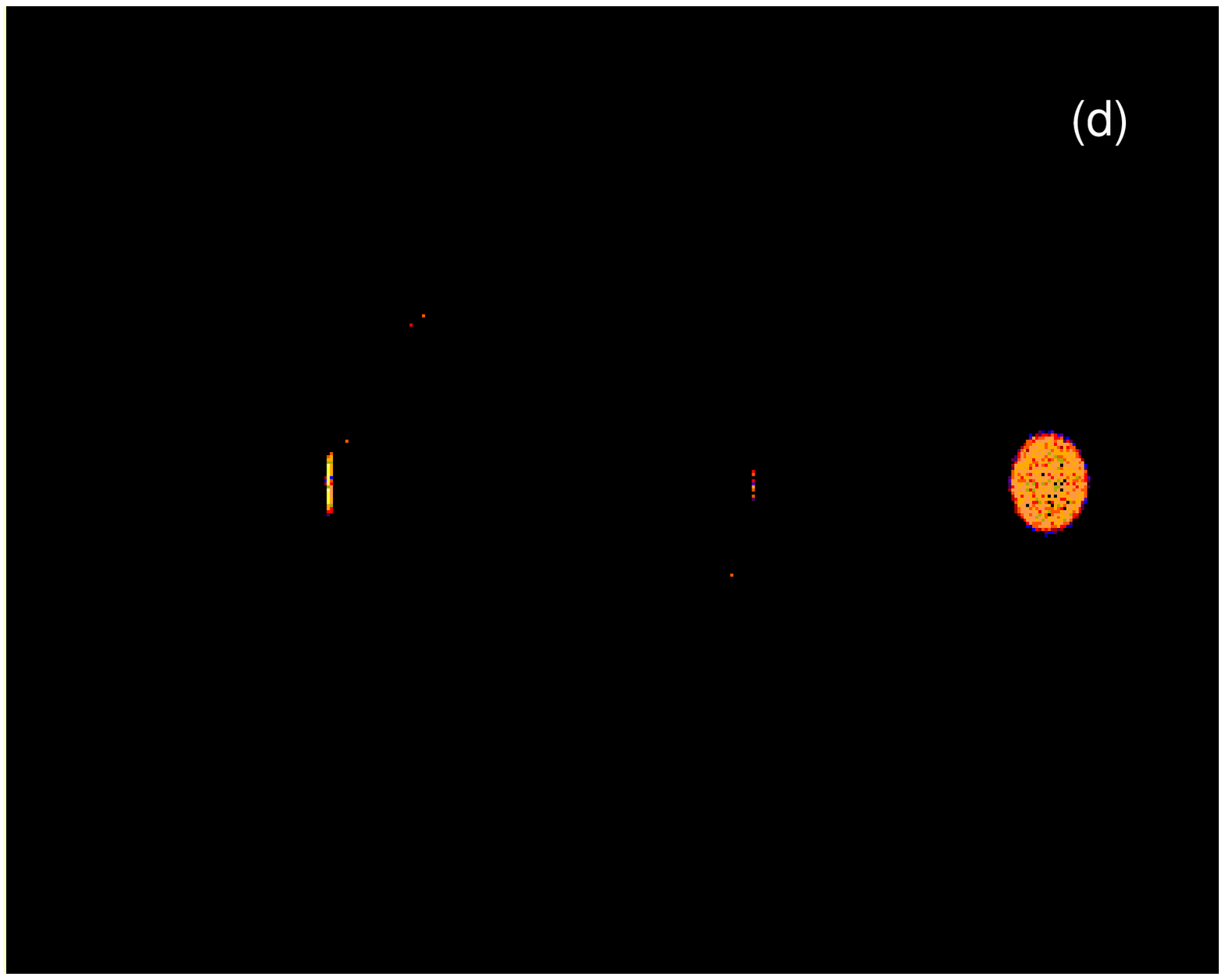}
\end{center}
\end{figure}

In Figure \ref{fig:movie_stills} we show four snapshots of the NS/BH
system as seen by an observer oriented edge-on to the orbital
plane. The binary masses are $m_1=10M_\odot$ and $m_2=1.4M_\odot$ and
the binary separation is $D=10M$. The two dominant effects here are
gravitational lensing and relativistic beaming. From weak lensing
theory, whenever there is co-alignment of the  source, lens, and
observer, an Einstein ring is formed, with high magnification. Lensing
by a black hole produces an infinite number of concentric Einstein
rings, each one at smaller radius and magnification. 

The primary Einstein ring can be seen in panel (a), when the NS is on
the far side of the BH relative to the observer. The NS then
progresses around its orbit to (b), when it reaches the point of
maximum blueshift and beaming towards the observer. Also seen in (b)
is the faint secondary Einstein ring, only appearing a quarter phase
later due to the time delay of photons orbiting around the BH before
reaching the observer. In panel (c) another Einstein ring is seen,
produced when the system is aligned in the order BH--NS--observer, and
the photons from the nearer NS are deflected $180^\circ$ by the BH on
the far side of the orbit before returning to the observer. By this
time, the NS has already moved roughly a quarter phase out of the way,
and thus does not appear co-linear with the BH. Finally, in panel (d)
the NS is at the point of maximum redshift, noticeably fainter than in
panel (b). 

\begin{figure}[h]
\caption{\label{fig:light_curve} {\it (top)} Normalized light curve corresponding
  to Figure \ref{fig:movie_stills}, with the time of each panel labeled
  accordingly. {\it (bottom)} Normalized light curve from the same
  system, but with binary separation $D=40M$.}
\begin{center}
\includegraphics[width=0.4\textwidth]{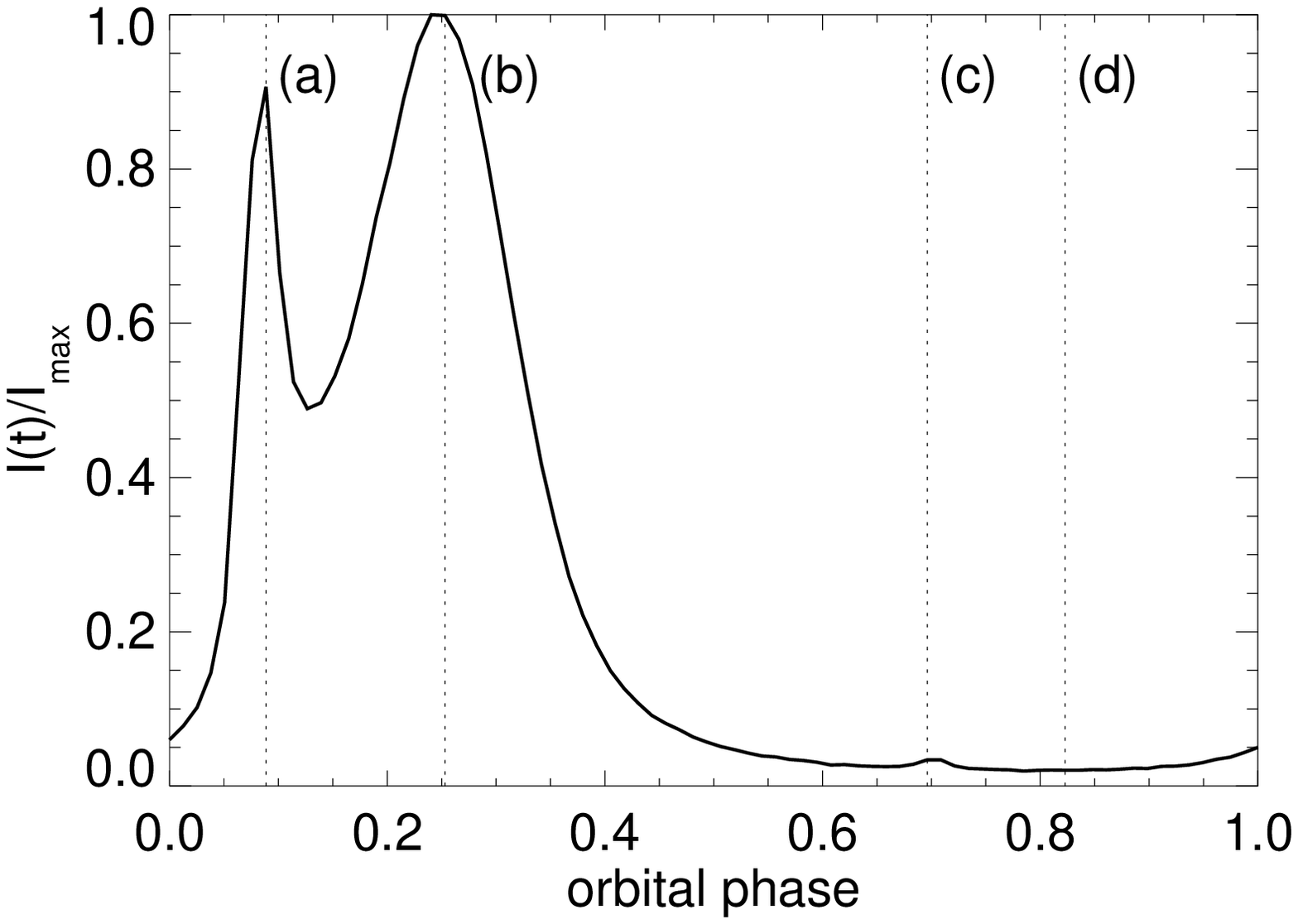}\\
\includegraphics[width=0.4\textwidth]{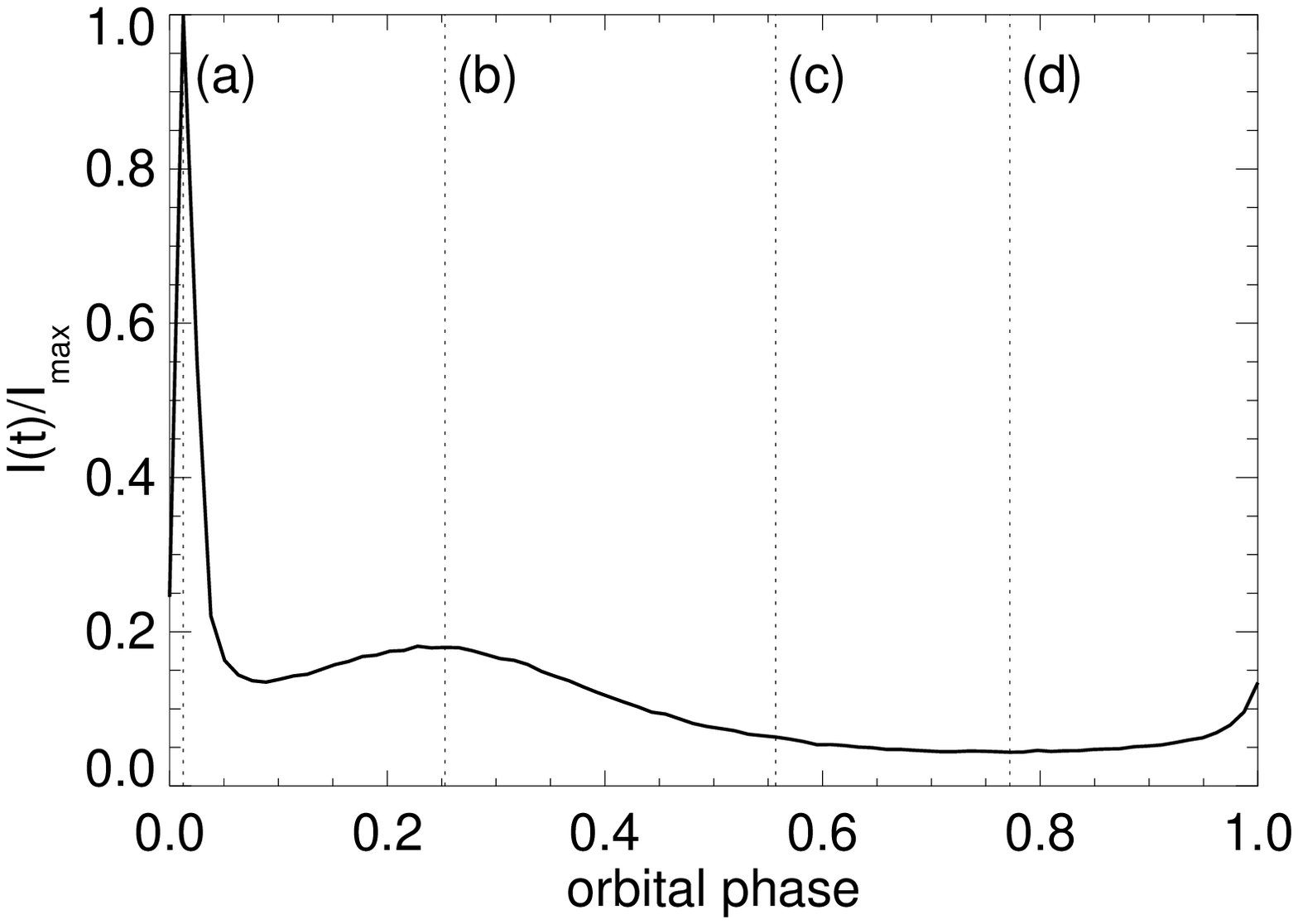}
\end{center}
\end{figure}

In Figure \ref{fig:light_curve} we show the bolometric light curve
corresponding to the snapshots in Figure \ref{fig:movie_stills}. At
this small binary separation of $10M$, the orbital velocity is roughly
a quarter of the
speed of light, so the dominant feature in the light curve is the
Doppler beaming as the NS moves towards and away from the
observer. The lensing magnification contributes a strong peak at (a),
and a much smaller peak visible at (c). In the bottom panel of Figure
\ref{fig:light_curve} we show the light curve from the same system at
binary separation $D=40M$. The orbital velocity, and thus relativistic
beaming effect, is much smaller, while the lensing still produces a
large magnification. The smaller orbital velocity also implies the
light-crossing time is shorter relative to the orbital period, so the
four phases corresponding to those shown in Figure 
\ref{fig:movie_stills} are more evenly spaced in time.

\begin{figure}[h]
\caption{\label{fig:lc_inc} Light curve dependence on observer
  inclination. The binary separation is $10M$ as in
  Fig.\ \ref{fig:movie_stills}. While the relativistic beaming effects
  are significant even at low inclinations, the lensing effects are
  only observable above $\sim 80^\circ$.}
\begin{center}
\includegraphics[width=0.8\textwidth]{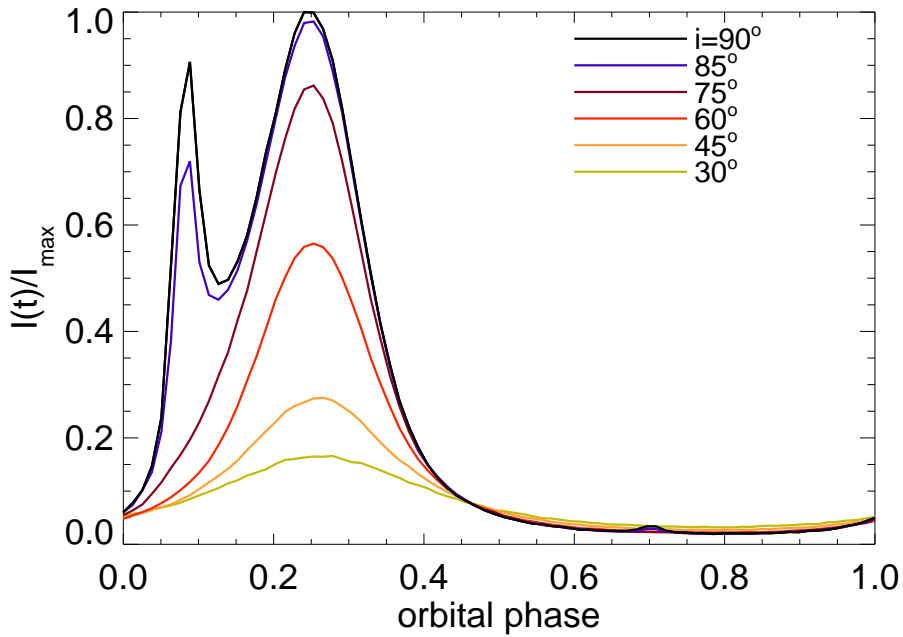}
\end{center}
\end{figure}

Figure \ref{fig:lc_inc} shows
the dependence of the light curve features on observer
inclination. Not surprisingly, the sharp lensing feature is only
visible at high inclination angles with $i\gtrsim 80^\circ$. Yet even
at $i=30^\circ$, the relativistic beaming is responsible for a significant
modulation in the light curve. This is because, at a given frequency,
the observed flux scales like the redshift cubed: $I_\nu \sim
(\nu_{\rm obs}/\nu_{\rm em})^3$. The bolometric flux modulation is greater by
another factor of the redshift. One simple way to understand this is
when the emitter has a blackbody temperature $T_{\rm em}$, it is
observed as a blackbody with temperature $T_{\rm obs}=T_{\rm
  em}(\nu_{\rm obs}/\nu_{\rm em})$, and of course the total flux
scales like $T^4$. For even the moderate inclination of $i=30^\circ$,
the ratio of peak-to-valley flux is $I_{\rm max}/I_{\rm min} \approx
3$, while for edge-on systems the modulation is an order of magnitude
greater.

\begin{figure}[h]
\caption{\label{fig:lc_energy} Light curve dependence on energy.
  The binary separation is $10M$ as in Fig.\ \ref{fig:movie_stills},
  and the observer inclination is $60^\circ$.}
\begin{center}
\includegraphics[width=0.8\textwidth]{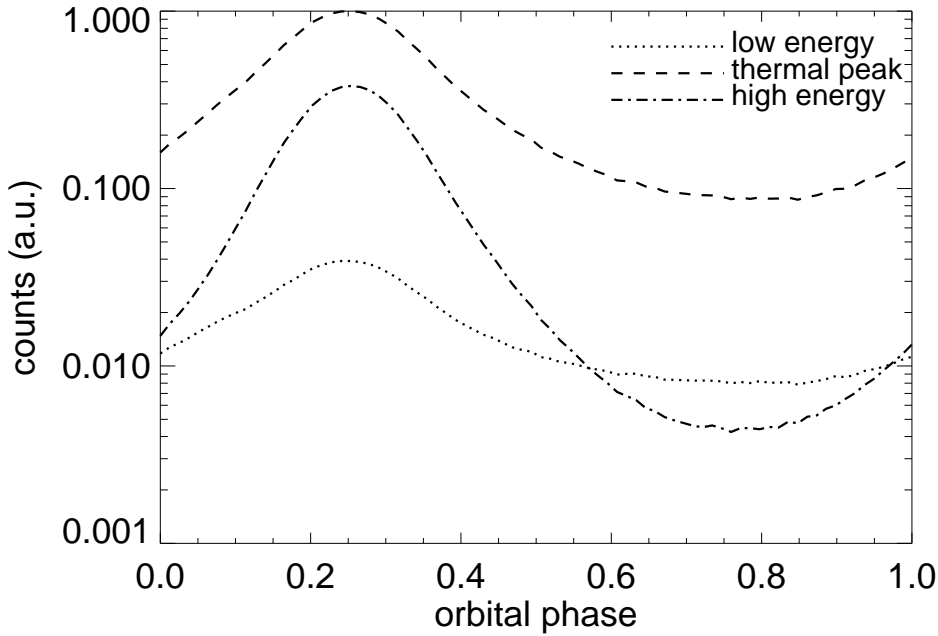}
\end{center}
\end{figure}

For a thermal emission spectrum, the magnitude of the Doppler
modulation also depends on the energy band at which it is observed. At
energies below the thermal peak, the slope of the spectrum is
increasing like $I_\nu \sim \nu^2$, so the net effect of beaming is
relatively modest: the observer sees a blue(red)-shifted portion of an
inherently fainter(brighter) part of the spectrum. The opposite is
true at energies above the thermal peak, where the relativistic beaming
combines with the inherent shape of the spectrum to enhance the effect
of Doppler modulation. These effects are shown in Figure
\ref{fig:lc_energy}, where we have divided the observed spectrum into
three broad bands, one covering the thermal peak, and one each above
and below. For an inclination angle of $60^\circ$, the bolometric
peak-to-valley amplitude is $\sim 10$, while the
high-energy flux is actually modulated by nearly a factor of 100. This
will have important implications for designing a search strategy for
these systems.

On the other hand, the gravitational lensing is achromatic, and generally
strongest when the source is moving transverse to the observer, so the
lensing modulation will have roughly the same effect on light curves
in different energy bands. 

\section{ELECTROMAGNETIC CHIRPS}

As shown in the previous section, the gamma-ray light curve from a
NS/BH binary is modulated at the orbital frequency with an amplitude that
is a function of the line-of-sight velocity. As the orbit shrinks due
to gravitational radiation losses, the frequency and amplitude of the
gamma-ray light curve increase. In other words, the EM light curve {\it
  chirps}, just like the GW signal. 

\begin{figure}[h]
\caption{\label{fig:inspiral} Electromagnetic modulation for an
  inspiraling NS/BH binary starting
  at orbital separation $50M$, with observer inclination angle $82^\circ$.
  The insets show zoomed-in views of the beginning and end of
  the inspiral.}
\begin{center}
\includegraphics[width=1.0\textwidth]{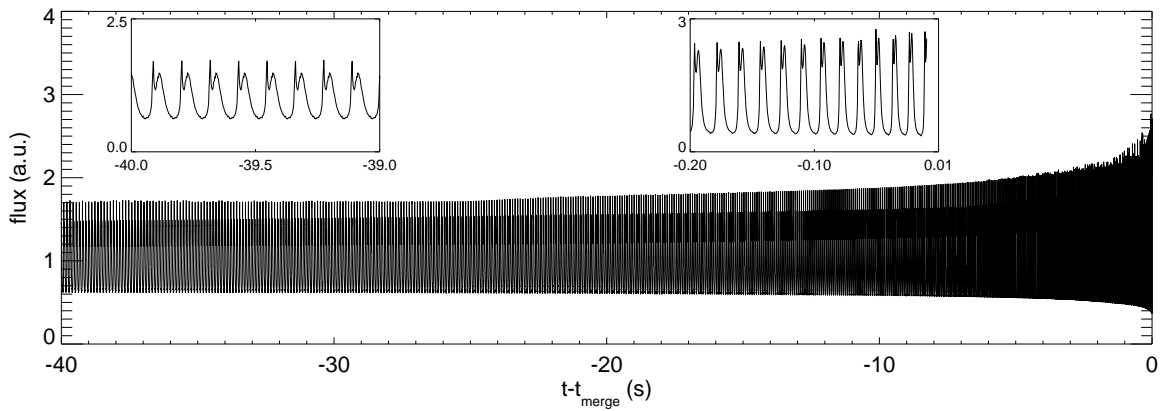}
\end{center}
\end{figure}

A NS/BH system with masses $1.4M_\odot$ and $10M_\odot$ will complete
roughly 700 orbits during the final minute before merger, entirely
within the LIGO band. A portion of the light curve corresponding to
this period is shown in Figure \ref{fig:inspiral}. To achieve
sufficiently high resolution with \pan over such a long time, we need
only calculate a handful of circular light curves, each at a different orbital
separation, and then interpolate between them according to the
inspiral evolution as governed by equation (\ref{eqn:Peters}). The
evolution is cut off at a separation $D=10M$, shortly before
merger. In practice, the light curve shown in Figure
\ref{fig:inspiral} should really be thought of as a {\it modulation window}
to be multiplied by the inherent flare luminosity, which may only last
a few seconds or less, depending on the emissivity model. 

At early times, when the orbital velocity is somewhat smaller, the
lensing peak of the light curve is larger than that due to the Doppler
beaming, but at late times the beaming dominates the modulation (see
Fig.\ \ref{fig:light_curve}).
As with traditional radial velocity observations of spectroscopic
binaries, the Doppler shift modulation will provide a degenerate
measurement of the BH mass and the binary inclination. For nearly
edge-on systems where the lensing peak is observable, the degeneracy
can be broken and an accurate BH mass can be determined, much like in
the case of transiting exoplanet systems. With a coincident
GW detection, the gravitational waveform will provide a different
degenerate measure of the binary masses and inclination. For extremely
precise light curve measurements, one could even imagine resolving the
width of the lensing peak, in turn giving information about the NS
radius and thus the equation of state. 

There is also the very real possibility that the source could be a
NS/NS binary, with either or both neutron stars initiating the
gamma-ray flare(s). In this case, the light curve would be simply
repeated with half the period, yielding two beaming and two lensing
peaks per orbit. Or if the flares occur at different times
(potentially due to different NS masses or magnetic field
configurations), the light curve would consist of two isolated flares,
modulated by the same chirping signature, but offset in phase by
$180^\circ$. 

\section{OBSERVATIONAL POTENTIAL}
To this point, it has not been necessary to specify the physical
mechanism that causes the gamma-ray flare on the NS surface. Anything
that leads to emission closely tied to the neutron star will give
qualitatively the same EM chirp as the light curves computed in the
previous sections. One promising model is that of resonant shattering
of the NS crust \citep{Tsang:2012,Tsang:2013}. In this scenario, the
tidal deformation of the NS, modulated by the orbital motion,
resonantly excites the crust-core interface eigenmode, leading to the
crust shattering and releasing  $\sim 10^{46-48}$ erg on an extremely
short time scale. Depending on the NS equation of state, this resonant
shattering flare takes place somewhere from 1 to 10 seconds before
merger, consistent with SGRB precursors seen with Swift
\citep{Troja:2010}. Another possible model that similarly deposits a
large amount of energy into the NS surface shortly before merger is
that of the unipolar inductor, where the accelerating magnetized NS
essentially forms an electric generator with the black hole or neutron  
star companion \citep{Hansen:2001,Mingarelli:2015,DOrazio:2016}. 

We propose to implement efficient searches for these characteristic chirp
signals in the data from existing time-domain observatories such as
Fermi in conjunction with any LIGO triggers. By using 
data analysis techniques similar to the matched filtering employed
by the LIGO collaboration, we hope to be able to extract long but weak
signals from a noisy background. Moreover, if the source parameters
are partially known from a coincident GW signal (in particular the
chirp mass and the merger time) then the parameters of the EM chirp
are strongly constrained, restricting the volume of parameter space
that must be explored and further reducing the false-alarm background
of the search, as well as its computational cost. Methods for
performing such a search in Fermi/GBM data are currently being
investigated and will be described in a companion paper (Dal Canton et
al.\ 2017). 

An important caveat in considering the effects of the EM chirp, in the
context of these models, is that high luminosities emitted thermally
from a small region, like the neutron star surface, will be optically
thick to pair-production ($\gamma + \gamma \rightarrow e^+ + e^-$),
resulting in a pair-photon fireball that only becomes optically thin
at surface of a much larger photosphere \citep{Goodman:1986}.

For photons with energies $E_\gamma \approx 500$ keV we can estimate
the optical depth to pair production (see e.g., \citet{Nakar:2005}):
\begin{equation}
\tau_{\gamma\gamma} \gtrsim 
\frac{L_\gamma\sigma_T}{4 \pi c E_\gamma R_{\rm NS}} \approx 10^{10} 
\frac{L_\gamma/(10^{46} {\rm erg\, s}^{-1})}{R_{\rm NS}/(10{\rm km})}
\, ,
\end{equation}
where $L_\gamma$ is the luminosity of photons near $E_\gamma$ and
$\sigma_T$ is the Thompson cross-section. Clearly, if all the energy
of these flares is deposited into photons above the pair-production
threshold, a relativistic pair-photon fireball is quickly formed,
which will expand relativistically until the pair-production freezes
out \citep{Goodman:1986}, resulting in a large photospheric radius
when compared to the orbital and gravitational scales. Additionally,
the photospheric surface is accelerated to large Lorentz factor
$\Gamma$, likely washing out any beaming effects from the
orbital dynamics.

For black-body emission, luminosity greater than $L \gtrsim 10^{42}$
erg s$^{-1}$, when confined to $R_{\rm NS} \approx 10$km ($T \gtrsim
20$keV), will have a sufficiently high-energy tail to be optically
thick to pair production. Thus, there exists an unfortunate trade-off
for potential EM chirp sources. Those brighter than $L \gtrsim
10^{42}$ erg s$^{-1}$, which are more easily detectable, will likely
be optically thick to pair-production and result in a pair-photon
fireball, potentially masking the EM chirp signature. Meanwhile, those
with luminosities below this threshold will be more difficult to
detect at extra-galactic distances. However, non-thermal emission
lacking significant contribution above the pair-production threshold
may avoid this limitation.

Resonant shattering flares, with luminosities of $\sim 10^{47-49}$ erg
s$^{-1}$, consistent with the observed SGRB precursors, are likely
pair-photon fireballs. However, emission from a black hole or neutron
star companion crossing the NS magnetic field lines in the unipolar
inductor model are significantly less energetic, with maximum
luminosity scaling as $L \sim 10^{40}$ erg
s$^{-1}$ $(B/10^{12} {\rm G})^2 (D/10^7 {\rm cm})^{-7}$, where $B$ is
the NS magnetic field and $D$ the binary separation \citep{Hansen:2001}.

Despite these potential complications, we are confident that the
basic EM chirp remains a robust and potentially very powerful
prediction for a counterpart to the GW signal from merging NS
binaries. In the near future, we encourage any wide-field survey with
sufficiently high time resolution to look for these characteristic
chirps preceding GW triggers. In the more distant future, we look
forward to using low-frequency GW observatories such as LISA to
identify precursor signals that would in turn give an early warning
for the time and sky location for ground-based events, triggering more
sensitive targeted EM observations \citep{Sesana:2016}.
\section*{Acknowledgments}

This work was partially supported by NASA grant ATP13-0077. TDC was
supported by an appointment to the NASA Postdoctoral Program at the
Goddard Space Flight Center, administered by Universities Space
Research Association under contract with NASA.

\newpage
\bibliography{em_chirp.bib}

\begin{thebibliography}{}
\expandafter\ifx\csname natexlab\endcsname\relax\def\natexlab#1{#1}\fi

\bibitem[{{Abbott} {et~al.}(2016{\natexlab{a}}){Abbott}, {Abbott}, {Abbott},
  {Abernathy}, {Acernese}, {Ackley}, {Adams}, {Adams}, {Addesso}, {Adhikari},
  \& et~al.}]{LIGO2}
{Abbott}, B.~P., {Abbott}, R., {Abbott}, T.~D., {et~al.} 2016{\natexlab{a}},
  Physical Review Letters, 116, 241103

\bibitem[{{Abbott} {et~al.}(2016{\natexlab{b}}){Abbott}, {Abbott}, {Abbott},
  {Abernathy}, {Acernese}, {Ackley}, {Adams}, {Adams}, {Addeso}, {Adhikari}, \&
  et~al.}]{LIGO1}
---. 2016{\natexlab{b}}, Physical Review Letters, 116, 061102

\bibitem[{Campanelli {et~al.}(2006)Campanelli, Lousto, \&
  Zlochower}]{Campanelli:2006}
Campanelli, M., Lousto, C.~O., \& Zlochower, Y. 2006, Physical Review D, 74,
  041501(R)

\bibitem[{{D'Orazio} {et~al.}(2016){D'Orazio}, {Levin}, {Murray}, \&
  {Price}}]{DOrazio:2016}
{D'Orazio}, D.~J., {Levin}, J., {Murray}, N.~W., \& {Price}, L. 2016, \prd, 94,
  023001

\bibitem[{{Eichler} {et~al.}(1989){Eichler}, {Livio}, {Piran}, \&
  {Schramm}}]{Eichler:1989}
{Eichler}, D., {Livio}, M., {Piran}, T., \& {Schramm}, D.~N. 1989, \nat, 340,
  126

\bibitem[{{Goodman}(1986)}]{Goodman:1986}
{Goodman}, J. 1986, \apjl, 308, L47

\bibitem[{{Hansen} \& {Lyutikov}(2001)}]{Hansen:2001}
{Hansen}, B.~M.~S., \& {Lyutikov}, M. 2001, \mnras, 322, 695

\bibitem[{Kelly {et~al.}(2007)Kelly, Tichy, Campanelli, \&
  Whiting}]{Kelly:2007}
Kelly, B.~J., Tichy, W., Campanelli, M., \& Whiting, B.~F. 2007, Physical
  Review D, 76, 024008

\bibitem[{{Mingarelli} {et~al.}(2015){Mingarelli}, {Levin}, \&
  {Lazio}}]{Mingarelli:2015}
{Mingarelli}, C.~M.~F., {Levin}, J., \& {Lazio}, T.~J.~W. 2015, \apjl, 814, L20

\bibitem[{{Nakar} {et~al.}(2005){Nakar}, {Piran}, \& {Sari}}]{Nakar:2005}
{Nakar}, E., {Piran}, T., \& {Sari}, R. 2005, \apj, 635, 516

\bibitem[{{Nissanke} {et~al.}(2010){Nissanke}, {Holz}, {Hughes}, {Dalal}, \&
  {Sievers}}]{Nissanke:2010}
{Nissanke}, S., {Holz}, D.~E., {Hughes}, S.~A., {Dalal}, N., \& {Sievers},
  J.~L. 2010, \apj, 725, 496

\bibitem[{{Peters}(1964)}]{Peters:1964}
{Peters}, P.~C. 1964, Physical Review, 136, 1224

\bibitem[{{Schnittman} \& {Krolik}(2013)}]{Schnittman:2013}
{Schnittman}, J.~D., \& {Krolik}, J.~H. 2013, \apj, 777, 11

\bibitem[{{Sesana}(2016)}]{Sesana:2016}
{Sesana}, A. 2016, Physical Review Letters, 116, 231102

\bibitem[{{Troja} {et~al.}(2010){Troja}, {Rosswog}, \& {Gehrels}}]{Troja:2010}
{Troja}, E., {Rosswog}, S., \& {Gehrels}, N. 2010, \apj, 723, 1711

\bibitem[{{Tsang}(2013)}]{Tsang:2013}
{Tsang}, D. 2013, \apj, 777, 103

\bibitem[{{Tsang} {et~al.}(2012){Tsang}, {Read}, {Hinderer}, {Piro}, \&
  {Bondarescu}}]{Tsang:2012}
{Tsang}, D., {Read}, J.~S., {Hinderer}, T., {Piro}, A.~L., \& {Bondarescu}, R.
  2012, Physical Review Letters, 108, 011102

\end{thebibliography}

\end{document}